\documentclass[a4paper]{jpconf}
\usepackage{graphicx}
\begin{document}
\title{Clustering in nuclear environment}

\author{ G. R\"opke}

\address{University of Rostock, Institut f\"ur Physik, 18051 Rostock, Germany}

\ead{gerd.roepke@uni-rostock.de}

\begin{abstract}
The properties of few-body clusters (mass number $A \le 4$) are modified if they are immersed in a nuclear medium.
In particular, Pauli blocking that reflects the antisymmetrization of the many-body wave function is responsible for the medium modification 
of light clusters and the dissolution with increasing density. A more consistent description is given with takes also the contribution of correlations in the continuum into account. The relation between cluster formation in warm dense matter and in nuclear structure is discussed.
\end{abstract}

\section{In-medium few-nucleon Schr\"odinger equation}

We consider the propagation of $A$ nucleons [$(A-Z)$ neutrons ($n$) and $Z$ protons ($p$)] in a nuclear medium 
described by the particle densities $n_n$ and $n_p$ as well as the temperature $T$. 
Using a thermodynamic Green function approach, the following in-medium wave equation can be derived (see \cite{R})
\begin{eqnarray}
&&[E_1^{\rm qu}(1)+\dots + E_1^{\rm qu}(A) - E^{\rm qu}_{A \nu}(P)]\psi_{A \nu P}(1\dots A)
\nonumber \\ &&
+\sum_{1'\dots A'}\sum_{i<j}[1- f(i)- f(j)]V(ij,i'j')\prod_{k \neq 
  i,j} \delta_{kk'}\psi_{A \nu P}(1'\dots A')=0\,.
\label{waveA}
\end{eqnarray}
This equation contains the effects of the medium in the quasiparticle shift $E_1^{\rm qu}(1)$
as well as in the Pauli blocking terms $ [1- f(i)- f(j)] $. 
The single-nucleon state $\{\vec p_1, \sigma_1,\tau_1\}$ is abbreviated by $i = 1$ etc.
The single-nucleon quasiparticle shift $\Delta E^{\rm SE}(1)=E_1^{\rm qu}(1)-\hbar^2 p_1^2/2 m$
is obtained from the single-nucleon self-energy which can be taken, for instance, in mean-field approximation.
Standard expressions give a rigid shift in the energy and, if the momentum dependence is taken into account, 
the replacement of the nucleon mass $m$ by an effective mass. The treatment of the single-nucleon shift
in solving Eq. (\ref{waveA}) is trivial \cite{R} and will not be investigated here. 
In particular, the rigid energy shifts can be absorbed by a shift $\Delta E^{\rm SE}_{A \nu}( P)=\Delta E^{\rm SE}(1)+\dots
+\Delta E^{\rm SE}(A)$ of the cluster quasiparticle energy 
\begin{equation}
E^{\rm qu}_{A \nu}( P)=\hbar^2 P^2/2 A m
+\Delta E^{\rm SE}_{A \nu}( P)+\Delta E^{\rm Pauli}_{A \nu}( P)+\Delta E^{\rm Coul}_{A \nu}( P)
\end{equation} 
and will not be considered here any more, the same also with the shift of the Coulomb energy.
Note that the in-medium few-nucleon Schr\"odinger equation (\ref{waveA}) describes the medium modification 
not only of possible bound states that may be considered as quasi-particles, but also the modification of the scattering states.

For exploratory calculations we use a simple separable interaction \cite{R2011}
with Gaussian form factor.
\begin{equation}
\label{sepa}
V({\bf p}_1, {\bf p}_2;{\bf p}'_1,{\bf p}'_2)=\lambda {\rm e}^{-({\bf p}_1-{\bf p}_2)^2/4 \gamma^2} {\rm e}^{-({\bf p}'_1-{\bf p}'_2)^2/4 
\gamma^2}\delta({\bf p}_1+{\bf p}_2-{\bf p}'_1-{\bf p}'_2)\,.
\end{equation}
The parameter values can be fixed by experimental data. In the case of the deuteron ($A=2$), 
with potential parameters $\lambda = -1287.4$ MeV  fm$^{3}$ and $\gamma = 1.474$ fm$^{-1}$ in Eq. (\ref{sepa}), 
the binding energy and rms radius of the free deuteron  are reproduced.

Our main concern are the effects of the Pauli blocking terms where the phase space occupation is described by the 
Fermi distribution function $ f(i)=[\exp(\hbar^2 p_i^2/2 m T-\mu_i/T)+1]^{-1}$. Here, the chemical potential $\mu_\tau$
is related to the nucleon density $n_{\tau_1}=(1/{\rm volume})\sum_{p_1,\sigma_1}f(1)$. In the case of zero temperature ($T=0$),
the Fermi function becomes the step function $\Theta(E^{\rm Fermi}_{\tau_1}-\hbar^2 p_1^2/2 m)$ where the chemical
potential coincides with the Fermi energy $E^{\rm Fermi}_{\tau}=(\hbar^2/2m)(3 \pi^2 n_\tau)^{2/3}$.
The Pauli blocking term leads to a shift in the energy $\Delta E^{\rm Pauli}_{A \nu}( P)$ \cite{R2011} and the in-medium modification 
of the wave function or the scattering phase shifts. The properties of bound states have been investigated elsewhere \cite{R}.
In particular we investigate the phase shifts 
and the consequences for the second virial coefficient. Different approaches are discussed.

\section{In-medium phase shifts for $A=2$}

We first consider the (deuteron like) case $A=2$. Furthermore
 we will use the Tamm-Dancoff (TD) expression  $ [1- f(i)- f(j)] \to  [1- f(i)][1- f(j)]$ 
 which neglects the hole-hole contributions that are of relevance in deriving the gap equation if pairing is considered. 
 
 For the separable potential, the T-matrix can be solved. With the integral
\begin{equation}
I(E) = -\frac{\lambda}{2\pi^2} 
%            \frac{m \gamma}{\hbar^2}
\int_0^\infty {\rm d}{p}\,
\frac{p^2}{\hbar^2 p^2/m-E}\,
{\rm e}^{-2 p^2/\gamma^2}\, [1-f( p)]^2
\end{equation} 
we can now calculate the scattering phase-shift depending on the energy $E$ of relative motion (zero angular momentum):
\begin{equation}
    \tan\left[\delta_0(E) \right] = 
      \frac{{\rm Im}\,{T(12,1'2',E)}}
           {{\rm Re}\,{T(12,1'2',E)}}
    =
      \frac{ -{\rm Im}\,{I(E)}}{1+{\rm Re}\,{I(E)}}
     \end{equation}
Note that for negative values of $E$, the solution of $1+{\rm Re}\,{I(-E_d)}=0$ gives the
in-medium binding energy, for instance, of the deuteron (isospin singlet channel). 

\begin{figure}[th] 
 	\includegraphics[width=0.6\textwidth]{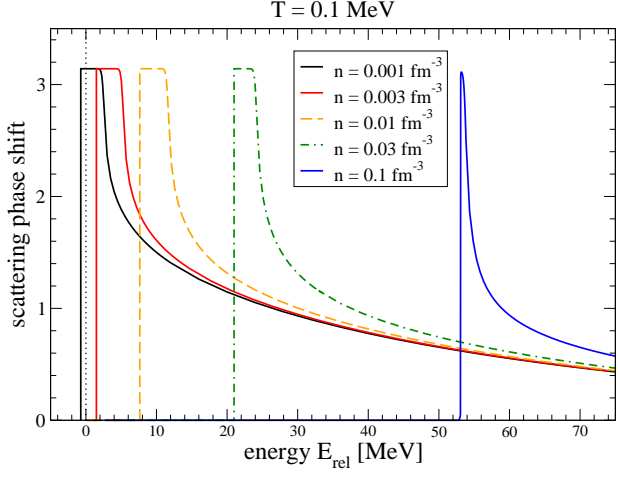}
	\caption{$T$ = 0.1 MeV scattering phase shifts for different densities, TD blocking.}
 \label{Fig.1} 
 \end{figure}

Obviously the bound state wave functions and energy eigenvalues $E_d(P,T,n_n,n_p)$ as
well as the scattering phase shifts become dependent on temperature
and density because of the Fermi distribution functions in the Pauli blocking term. 
We focus on the  scattering phase shifts $\delta_0(E;P,T,n_n,n_p)$. For $P=0$ 
and $n_n=n_p=n/2$, some results are shown in Fig.~\ref{Fig.1}.

 \begin{figure}[th] 
	\includegraphics[width=0.6\textwidth]{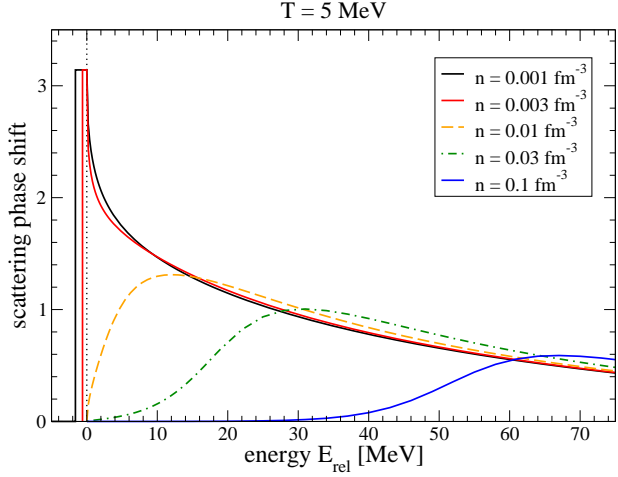}
	\caption{$T$ = 5 MeV scattering phase shifts for different densities, TD blocking.}
 \label{Fig.2} 
 \end{figure}  

Having the scattering phase shifts to our disposal, we can calculate the second virial coefficient
according to the generalized Beth-Uhlenbeck formula \cite{SRS}. In addition to the bound state contribution  
$b^{\rm bound}_{^3S_1}(0,T,n_n,n_p)=3 [\exp (E_d(0,T,n_n,n_p)/T)-1]$ ($P$ dependence of the in-medium binding energy neglected)  
we have the contribution of scattering states,
\begin{equation}
b^{\rm sc}_{^3S_1}(P,T,n_n,n_p)=\frac{3}{\pi T}\int_0^\infty {\rm d}E\, {\rm e}^{-E/T}\left[\delta_0(E;P,T,n_n,n_p)-\frac{1}{2}
\sin\left[2 \delta_0(E;P,T,n_n,n_p)\right]\right]
\end{equation}
If there exist a bound state (below the critical density $n_{\rm cr}(T)$ where, e.g., $n_{\rm cr}(0.1\, {\rm MeV})=0.0016167$ fm$^{-3}$, 
$n_{\rm cr}(5\, {\rm MeV})=0.005472$ fm$^{-3}$), it is possible to combine the bound state and the scattering state contribution to the second virial coefficient 
extending the scattering phase shift to the bound state energy, see Fig.~\ref{Fig.1}. 
There is a smooth behavior near the Mott point where the bound state is dissolved because of Pauli blocking.
We give also a simple interpolation formula that can be used to estimate the contribution of the continuum correlations to the second
virial coefficient and, furthermore, to the density fraction of correlated states:
\begin{equation}
b^{\rm sc}_{^3S_1}\approx \frac{3 [0.008709 \exp(-0.15403 T) [0.3007+0.6662 \exp(-0.1014 T)] +n^2]}{0.008709 \exp(-0.15403 T)
+n^2/[-0.1012+0.08611 \exp(0.09432 T)] \exp[(26.3+483.8/T)n]}
\end{equation}
%\begin{equation}
%b^{\rm sc}_{^3S_1}(P,T,n_n,n_p)\approx 3\frac{0.008709 \exp(-0.15403 T) [0.3007+0.6662 \exp(-0.1014 T)] +n^2}{0.008709 \exp(-0.15403 T)
%+n^2/[-0.1012+0.08611 \exp(0.09432 T)] \exp[(26.3+483.8/T)n]}
%\end{equation}
for $T > 1$ MeV. At lower temperatures, $b^{\rm sc}_{^3S_1}\propto \exp[-(\hbar^2/mT) (3 \pi^2 n/2)^{2/3}]$.

An important feature is that at very low temperatures the phase shifts are constant,  $\delta = \pi$, not only up to the edge of the continuum
of scattering states but up to the Fermi energy (at $T=0$, energy conserving nucleon-nucleon scattering 
for energies below $E_\tau^{\rm Fermi}$ is not possible because of Pauli blocking). 
Thus, the bound states merge with the continuum of scattering states at zero temperature only if the bound state energy per nucleon 
becomes larger than the Fermi energy. 

At higher temperatures these sharp structures in the phase shifts shown in Fig.~\ref{Fig.1} are washed out, see Fig.~\ref{Fig.2}. 
The bound state will be dissolved already when it merges with the continuum of scattering states 
($E=0$ when the self-energy rigid shifts are absorbed by the chemical potential). 
%In the approach discussed here where the self-energy terms are considered as rigid shift and can be absorbed by the chemical potential, 
%the edge of the continuum is at $E=0$.

%The contribution of scattering states to the second virial coefficient can be approximated by the following expression
%\begin{equation}
%b^{\rm sc}_{^3S_1}\approx \frac{3 [0.008709 \exp(-0.15403 T) [0.3007+0.6662 \exp(-0.1014 T)] +n^2]}{0.008709 \exp(-0.15403 T)
%+n^2/[-0.1012+0.08611 \exp(0.09432 T)] \exp[(26.3+483.8/T)n]}
%\end{equation}
%%\begin{equation}
%%b^{\rm sc}_{^3S_1}(P,T,n_n,n_p)\approx 3\frac{0.008709 \exp(-0.15403 T) [0.3007+0.6662 \exp(-0.1014 T)] +n^2}{0.008709 \exp(-0.15403 T)
%%+n^2/[-0.1012+0.08611 \exp(0.09432 T)] \exp[(26.3+483.8/T)n]}
%%\end{equation}
%for $T > 1$ MeV. At lower temperatures, $b^{\rm sc}_{^3S_1}\propto \exp[-(\hbar^2/mT) (3 \pi^2 n/2)^{2/3}]$. 

\section{Particle-particle RPA and pairing  for $A=2$}
\begin{figure}[h]
\begin{minipage}{20pc}
\includegraphics[width=20pc]{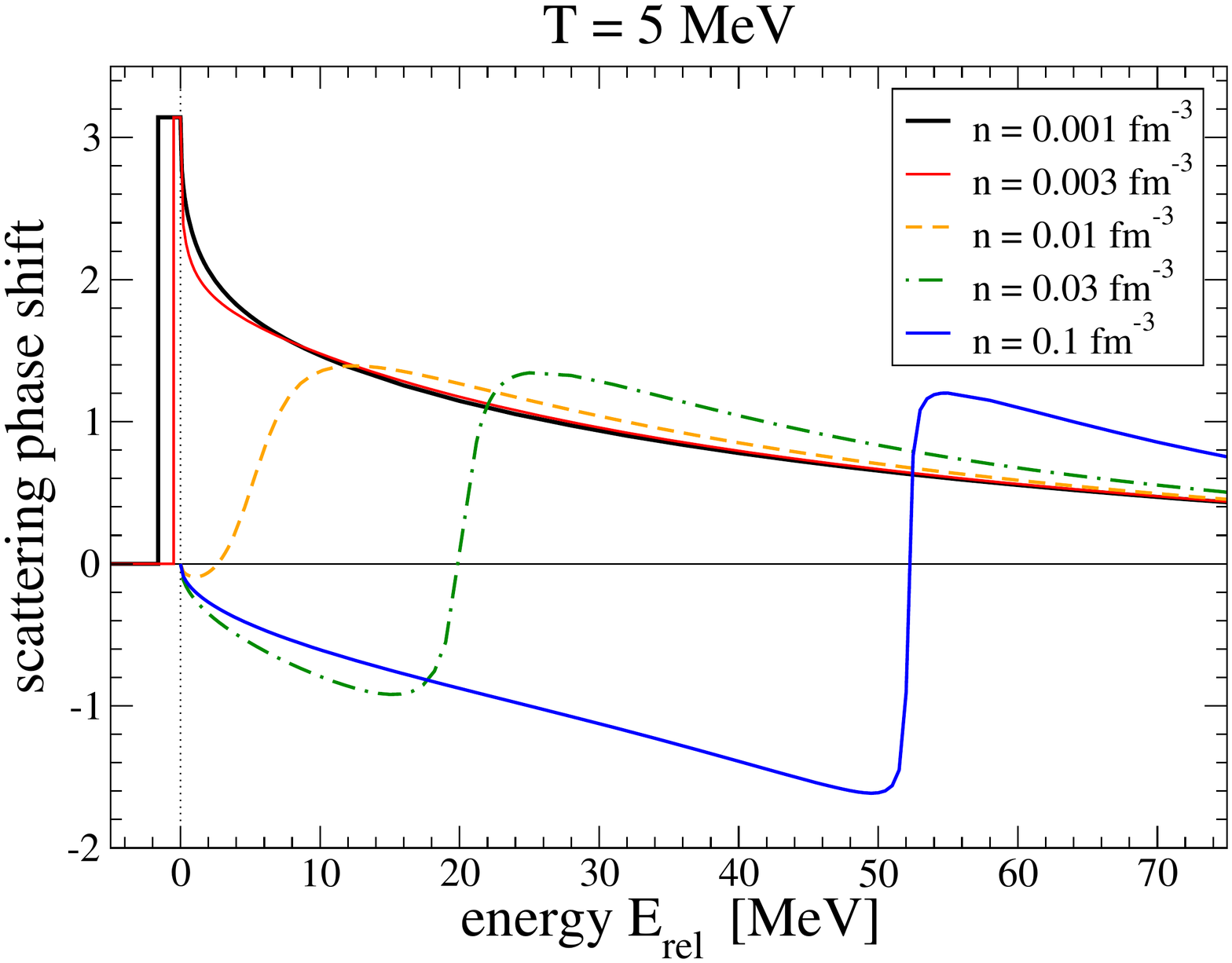}
\caption{$T$ = 5 MeV; ppRPA blocking.}
\label{Fig.3}
\end{minipage}\hspace{0pc}%
\begin{minipage}{20pc}
\includegraphics[width=20pc]{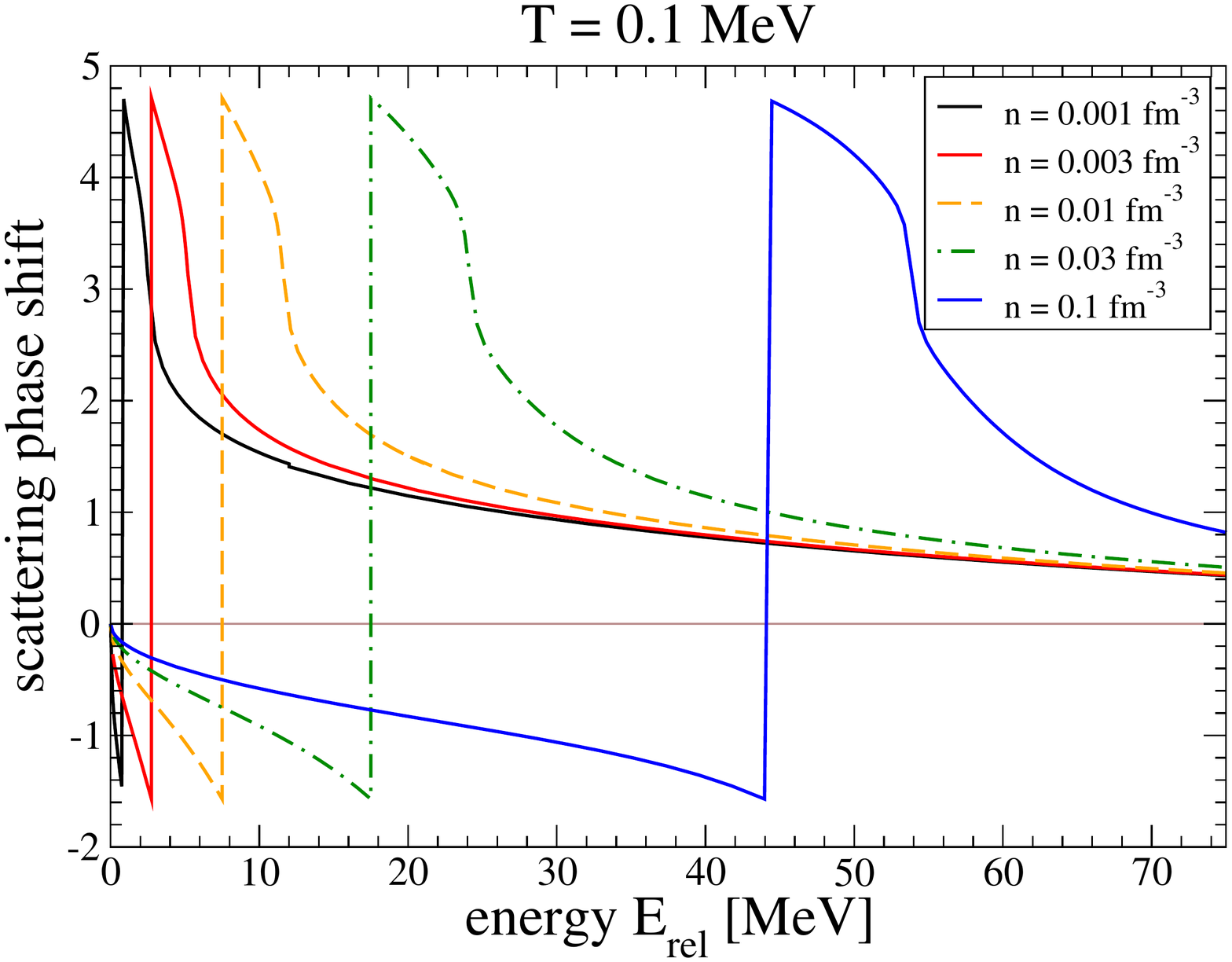}
\caption{$T$ = 0.1 MeV; ppRPA blocking.}
\label{Fig.4}
\end{minipage} 
\end{figure}

Coming back to the so-called particle-particle Random-Phase Approximation  (ppRPA) \cite{RS},
the Pauli blocking term reads  $ [1- f(i)- f(j)] $ 
 which takes the hole-hole contributions into account. The in-medium phase shifts are shown in Fig.~\ref{Fig.3} for
 moderate temperatures of warm dense matter.   At very low temperatures, sharp structures arise similar to the Tamm-Dancoff case,
 see Fig.~\ref{Fig.4}.
%  \begin{figure}[th] 
%  	\includegraphics[width=0.6\textwidth,angle=0]{phshiftsPP5.pdf}
% 	\caption{$T$ = 5 MeV scattering phase shifts for different densities, ppRPA blocking.}
%  \label{Fig.3} 
%  \end{figure}  
% 
% \begin{figure}[th] 
%  	\includegraphics[width=0.6\textwidth]{phshiftsPP01.pdf}
% 	\caption{$T$ = 0.1 MeV scattering phase shifts for different densities, ppRPA blocking.}
%  \label{Fig.4}
%  \end{figure}  
As in the Tamm-Dancoff case, we obtain a jump of the phase shifts at energies that are near to the 
 energies of bound state formation, but there is also a general negative shift what resembles the effect of repulsion.

 We will not discuss here the consequences  due to the use of different expressions for the Pauli blocking term  (the critical densities 
 are now $n_{\rm cr}(0.1\, {\rm MeV})=0.000235476$ fm$^{-3}$, 
$n_{\rm cr}(5\, {\rm MeV})=0.0045349$ fm$^{-3}$). 
 The difference is the term $f(i)f(j)\propto n^2$ that is of higher order in the density, and one may argue that there are also further terms 
 missing if we consider this order of density. However, there is a more fundamental difference between both expressions for the 
 Pauli blocking. 
 The Tamm-Dancoff expression is used, for instance, in the Brueckner theory to calculate the quasiparticle energy shifts of the nucleons.
 On the other hand, pairing may occur in nuclear matter at low temperatures. The corresponding Gorkov equation follows only if the ppRPA
 expression for the Pauli blocking is used.
 A detailed discussion is given in Ref. \cite{SRS}.

\section{$\alpha$-particle case $(A=4)$}

Because of the relatively high binding energy (28.3 MeV), the $\alpha$-particle is of special interest in describing correlations and cluster formation 
in warm dense matter. The wave function of the free $\alpha$-particle can be approximated solving the four-particle Schr\"odinger equation with a
pair potential (\ref{sepa}), where the potential parameters $\lambda = -1449.6$ MeV  fm$^{3}$ and $\gamma = 1.152$ fm$^{-1}$ 
are determined by the binding energy and the rms radius of the  $\alpha$-particle.

\begin{figure}[th] 
 	\includegraphics[width=0.6\textwidth]{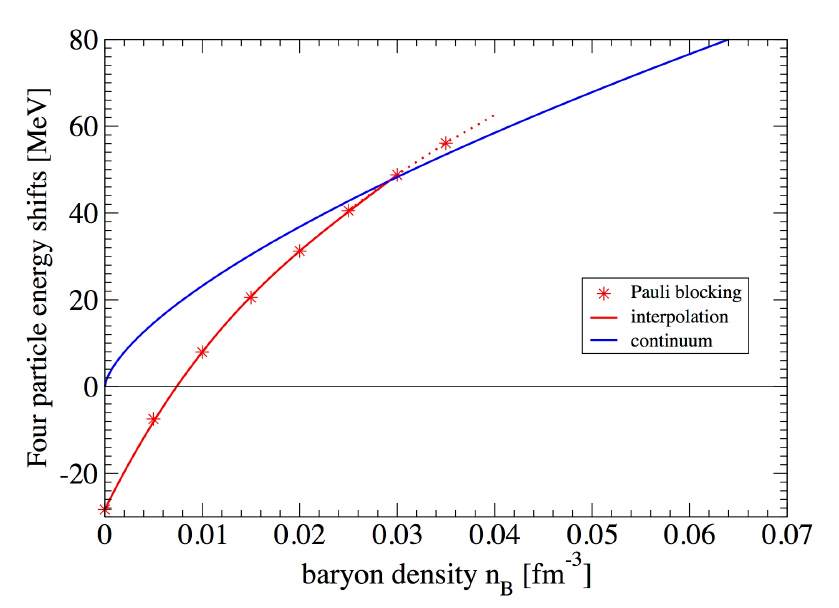}
	\caption{Internal four-nucleon energy (no c.o.m. motion) in a medium with nucleon density $n=n_n+n_p$. 
	The continuum edge of free single-particle states is given by $4 E_{\rm Fermi}$. 
	At zero baryon density, the four-nucleon energy is given by the binding energy of the $\alpha$ particle, 
	$B^0_\alpha =28.3$ MeV. With increasing density, the binding energy $B^0_\alpha$ is 
	decreasing due to the Pauli blocking. The four-nucleon bound state disappears at
	$n_B \approx 0.03$ fm$^{-3}$. A fit to the calculated values is given in Ref. \cite{Po}.}
 \label{Fig:4nuc} 
 \end{figure}

The medium modifications of the bound state energy and of the continuum states are of interest. Within a variational approach,
results are obtained as described elsewhere \cite{Po}. As example, we give here the shift of the binding energy of the $\alpha$-particle,
see Fig.~\ref {Fig:4nuc}.

Using a variational approach, we considered the shift of the $\alpha$ particle in zero temperature matter \cite{Po}. 
Whereas at finite temperatures the  bound state becomes dissolved already when the binding energy becomes zero,
at zero temperature the bound state can disintegrate only when the free particle states are not Pauli blocked.
Above the Fermi energy the decay is possible. This way we can understand why at finite temperatures the Mott point is given
by the zero binding energy condition, whereas at zero temperature the Fermi energy is of relevance. 
We conclude that at zero temperature the $\alpha$-like bound state is not dissolved at $n\approx 0.01$ fm$^{-3}$ where it merges 
in the continuum of unbound scattering states but at $n\approx 0.03$ fm$^{-3}$ where it can decay into unbound, free states above the 
Fermi energy because they are not occupied. This value $n\approx 0.03$ fm$^{-3}$ is in agreement with other approaches, in particular the fully antisymmetrized THSR  formalism, see \cite{Tohsaki}.

It should be mentioned that we may calculate the virial coefficient for the four-nucleon case by rescaling the two-nucleon result.
Another important issue is the treatment of correlations in the medium. 
Already the ppRPA expression for the Pauli blocking leads to the formation of Cooper pairs in the medium. The THSR ansatz  allows to 
treat four-nucleon correlations also in the medium, taking into account full antisymmetrization.\\

\section*{Acknowledgments}

The author acknowledges the contributions of P. Schuck, Y. Funaki, H. Horiuchi, Zhongzhou Ren,  A. Tohsaki, Chang Xu, T. Yamada,
and Bo Zhou to this work.


\section*{References}


\begin{thebibliography}{9}


\bibitem{R}
R\"opke G 2009 {\it Phys. Rev.} C {\bf 79} 014002


\bibitem{R2011}
  R\"opke G 2011
  {\it Nucl. Phys.} A {\bf 867} 66  
  
\bibitem{RS}  
   Ring P and Schuck P 1980 {\it The Nuclear Many-Body Problem} (Berlin: Springer)

\bibitem{SRS}
Schmidt M {\it et al.} 1990
 {\it Ann.\ Phys.} {\bf 202} 57 

 \bibitem{Po}
 R\"opke G {\it et al.} 2014 Bound clusters on top of doubly magic nuclei {\it Preprint} arXiv:1407.0510 [nucl.theory]
 
 \bibitem{Tohsaki} 
 Hiroki Takemoto {\it et al.} 2004 {\it Phys. Rev.} C {\bf 69} 035802
 
\end{thebibliography}
\end{document}